\documentclass[conference]{IEEEtran}
\usepackage{cite}
\usepackage{amsmath,amssymb,amsfonts}
\usepackage{algorithmic}
\usepackage{graphicx}
\usepackage{textcomp}
\usepackage{xcolor}
\usepackage{comment}
\usepackage{enumitem}
\usepackage{tikz}
\usetikzlibrary{shapes, arrows.meta, positioning, calc, decorations.pathreplacing}
\usepackage[margin=1in]{geometry}
\usepackage{booktabs}
\usepackage{multirow}
\usepackage{hyperref}
\usepackage{makecell}
\usepackage{siunitx}
\def\BibTeX{{\rm B\kern-.05em{\sc i\kern-.025em b}\kern-.08em
    T\kern-.1667em\lower.7ex\hbox{E}\kern-.125emX}}

\begin{document}

\title{Latency and Cost of Multi-Agent Intelligent Tutoring at Scale}

\author{\IEEEauthorblockN{Iizalaarab Elhaimeur}
\IEEEauthorblockA{\textit{Center for Real-Time Computing} \\
\textit{Computer Science Department}\\
\textit{Old Dominion University}\\
Norfolk, VA \\
ielha003@odu.edu}
\and
\IEEEauthorblockN{Nikos Chrisochoides}
\IEEEauthorblockA{\textit{Center for Real-Time Computing} \\
\textit{Computer Science and Physics Departments}\\
\textit{Old Dominion University}\\
Norfolk, VA \\
nikos@cs.odu.edu}
}

\maketitle

\begin{abstract}
Multi-agent LLM tutoring systems improve response quality through agent specialization, but each student query triggers several concurrent API calls whose latencies compound through a parallel-phase maximum effect that single-agent systems do not face. We instrument ITAS, a four-agent tutoring system built on Gemini 2.5 Flash and Google Vertex AI, across three throughput tiers (Standard PayGo, Priority PayGo, and Provisioned Throughput) and eleven concurrency levels up to 50 simultaneous users, producing over 3,000 requests drawn from a live graduate STEM deployment. Priority PayGo maintains flat sub-4-second response times across the full load range; Standard PayGo degrades substantially under classroom-scale concurrency; and Provisioned Throughput delivers the lowest latency at low concurrency but saturates its reserved capacity above approximately 20 concurrent users. Cost analysis places both pay-per-token tiers well below the price of a STEM textbook per student per semester under a worst-case usage ceiling. Provisioned Throughput, expensive under continuous provisioning, becomes cost-competitive for institutions that can predict and concentrate their traffic toward high utilization. These results provide concrete tier-selection guidance across deployment scales from a single seminar to a university-wide rollout.
\end{abstract}


\section{Introduction}

Large language models are entering education as tutoring assistants, but deployment has outpaced understanding of latency at scale. A student who waits 4 seconds for a response will find the interaction natural; a student who waits 10 seconds during a lab session with 39 classmates will stop asking questions. This constraint becomes acute when the tutoring system uses a multi-agent architecture, where a single student query triggers not one but several LLM calls, because latency compounds in ways that single-agent chatbots do not experience.

Multi-agent architectures are increasingly favored in educational AI because agent specialization improves response quality~\cite{elhaimeur2025toward}: a dedicated code-debugging agent produces better debugging advice than a general-purpose prompt, and a video-context agent that processes lecture content provides more relevant pedagogical guidance than one that lacks it. ITAS, the Intelligent Tutoring Agent System we developed for university STEM courses, uses three parallel specialist agents (video, code, and guidance) followed by a sequential synthesizer that merges their outputs into a coherent response. This spoke-and-wheel design yields four Gemini 2.5 Flash API calls per student interaction, each of which must complete before the student sees a response.

Cloud inference providers offer multiple throughput tiers that trade cost for performance guarantees. Google Vertex AI~\cite{vertexai} provides three options relevant to educational workloads: \emph{Standard PayGo}, which charges per token on a shared inference pool; \emph{Priority PayGo}~\cite{prioritypaygo}, which charges approximately 1.8$\times$ the standard rate but places requests in a priority queue on a global endpoint; and \emph{Provisioned Throughput}~\cite{provisionedthroughput}, which reserves dedicated capacity measured in Generative AI Scale Units (GSUs) at a fixed monthly rate. The latency implications of these tiers for multi-agent pipelines, where a single user interaction generates four concurrent API calls, have not been quantified.

This paper addresses two research questions:

\textbf{RQ1:} How do different throughput tiers affect latency in multi-agent LLM pipelines, and which tier is optimal for classroom-scale deployment?

\textbf{RQ2:} Can multi-agent tutoring scale from a small seminar to a university-wide deployment of 25,000 students, and at what cost?

We answer these questions through controlled benchmarks replaying approximately 100 real student queries from a live deployment of ITAS in a graduate STEM seminar. These queries are replayed at 11 concurrency levels (1 through 50 simultaneous users) across all three Vertex AI throughput tiers, producing over 3,000 instrumented requests with per-agent timing, token counts, and reliability data.

Our contributions are:
\begin{enumerate}[leftmargin=*]
    \item \textbf{Latency model.} Formalization of multi-agent LLM pipeline latency as a maximum-of-parallel distribution, showing that variance minimization, not mean minimization, is the key optimization target.
    \item \textbf{Empirical evaluation.} Three-tier benchmark across 11 concurrency levels (1--50 users) and 3,000+ real instrumented requests, identifying Priority PayGo as the best-performing tier across evaluated regimes.
    \item \textbf{Crossover analysis.} Identification of a provisioned--priority crossover at $c \approx 20$, with characterization of the cost and latency tradeoffs on either side.
    \item \textbf{Cost model.} Per-student deployment cost analysis from seminar to university scale, demonstrating sub-textbook cost under worst-case usage and a favorable cost structure for institutions with predictable traffic.
\end{enumerate}

\section{Related Work}

\subsection{LLM Inference Optimization}

Efficient LLM serving has been an active systems research area since the
emergence of large autoregressive models. Yu~et~al.\ introduced iteration-level
scheduling (Orca~\cite{yu2022orca}), eliminating padding waste and achieving
36$\times$ throughput improvement over static batching. Kwon~et~al.\ proposed
PagedAttention (vLLM~\cite{kwon2023vllm}), reducing KV-cache memory
fragmentation and improving throughput by 2--4$\times$. On the latency side,
speculative decoding~\cite{leviathan2023speculative,chen2023speculative} uses
a small draft model to propose tokens verified in parallel by the target model,
yielding 2--3$\times$ end-to-end speedup. More recent work on prefill-decode
disaggregation, Sarathi-Serve~\cite{agrawal2024sarathi} and
DistServe~\cite{zhong2024distserve}, separates compute-bound prefill from
memory-bound decode phases onto distinct hardware, reducing tail latency under
SLO constraints by up to 5.6$\times$. At the cost level,
FrugalGPT~\cite{chen2024frugalgpt} cascades models of increasing capability
to reduce spend by up to 98\%, and Chen~et~al.~\cite{chen2024scalinglaws}
derive scaling laws for compound LLM inference systems showing that accuracy
can decrease non-monotonically with the number of calls. Recent industry
benchmarks and provider reports (e.g., Artificial Analysis, MLCommons
Inference) document single-call latency variability across cloud endpoints,
but do not extend to multi-call pipelines or characterize how tail-latency
distributions compound across concurrent agent invocations.

All of these works target \emph{single-call or single-model} workloads and
optimize either mean latency or aggregate throughput. None model the compound
tail-latency distribution that arises when multiple heterogeneous agents execute
in parallel and the pipeline must wait for the slowest to complete; the
parallel-phase maximum effect we characterize is structurally absent from
existing serving literature.

\subsection{Multi-Agent LLM Systems}

Multi-agent LLM frameworks have grown rapidly since the ReAct
paradigm~\cite{yao2023react} demonstrated that interleaving chain-of-thought
reasoning with tool calls substantially improves task performance.
AutoGen~\cite{wu2023autogen} generalizes this to configurable multi-agent
conversations, enabling human, LLM, and tool-backed agents to collaborate on
complex tasks. MetaGPT~\cite{hong2024metagpt} encodes standardized operating
procedures into role-based agent collaboration, reducing cascading
hallucination errors in software generation. Du~et~al.~\cite{du2024improving}
show that multi-round debate between LLM instances improves factuality and
mathematical reasoning, while Guo~et~al.~\cite{guo2024multiagent} survey
progress and challenges across the field. A common finding is that agent
specialization, assigning distinct roles with focused prompts, improves both
output quality and reliability relative to a single monolithic agent.

Latency and cost under concurrent production load receive little attention in
this literature. Frameworks report task-completion quality and communication
overhead but do not characterize how response time scales with the number of
simultaneous users, nor do they model the variance amplification that parallel
fan-out introduces. Despite improvements in reasoning quality, the latency
behavior of multi-agent pipelines under concurrent deployment remains
uncharacterized, a gap our work fills with a controlled concurrency sweep
across three throughput tiers.

\subsection{ITS Deployment and LLMs in Education}

Intelligent tutoring research has long demonstrated that individualized
instruction improves learning outcomes~\cite{bloom1984twosigma,vanlehn2011relative},
and rule-based cognitive tutors established that automated instruction can
approach the effectiveness of one-to-one human tutoring at scale~\cite{anderson1995cognitive}.
LLMs have renewed this ambition. Tack and Piech~\cite{tack2022aiteacher} and
the BEA 2023 shared task~\cite{tack2023bea} revealed measurable gaps between
LLM and human pedagogy, while more recent work shows the gap closing:
Vanzo~et~al.~\cite{vanzo2024gpt4homework} found GPT-4 homework tutoring
improved learning outcomes in a randomized controlled trial, and Google
DeepMind's LearnLM~\cite{jurenka2024learnlm} demonstrated that
pedagogically fine-tuned Gemini models are consistently preferred over
prompted baselines by teachers and learners. Kasneci~et~al.~\cite{kasneci2023chatgpt}
survey the broader landscape of LLM opportunities in education.

The systems dimension of LLM tutoring remains largely unaddressed. A
systematic scoping review of 118 papers by Yan~et~al.~\cite{yan2024practical}
explicitly identifies cost and scalability as open barriers to adoption; none
of the surveyed works provides quantitative latency or per-student cost
analysis under realistic concurrent classroom load. This leaves open the
question of whether LLM-based tutoring systems can scale economically and
interactively in real classrooms, the question our paper answers.

\section{System Architecture}

\subsection{Multi-Agent Pipeline}

\begin{figure}[t]
\centering
\begin{tikzpicture}[
    scale=0.65,
    transform shape,
    node distance=1.1cm and 1.4cm,
    every node/.style={font=\small},
    block/.style={rectangle, draw, rounded corners=3pt, minimum height=0.75cm, minimum width=2.0cm, align=center, fill=#1!12, draw=#1!60, line width=0.6pt},
    block/.default=blue,
    arrow/.style={-{Stealth[length=5pt]}, thick, color=#1!70},
    arrow/.default=black,
    timing/.style={font=\scriptsize\itshape, color=gray!80},
]

\node[block=gray, minimum width=2.8cm] (query) {Student Query};

\node[block=red, below left=1.1cm and 1.6cm of query] (video) {Video Agent};
\node[block=blue, below=1.1cm of query] (guidance) {Guidance Agent};
\node[block=green, below right=1.1cm and 1.6cm of query] (code) {Code Agent};

\node[block=purple, below=1.1cm of guidance, minimum width=2.8cm] (synth) {Synthesizer Agent};

\node[block=gray, below=1.0cm of synth, minimum width=2.8cm] (response) {Student Response};

\draw[arrow=gray] (query.south) -- ++(0,-0.25) -| (video.north);
\draw[arrow=gray] (query.south) -- (guidance.north);
\draw[arrow=gray] (query.south) -- ++(0,-0.25) -| (code.north);

\draw[arrow=red] (video.south) |- (synth.west);
\draw[arrow=blue] (guidance.south) -- (synth.north);
\draw[arrow=green] (code.south) |- (synth.east);

\draw[arrow=purple] (synth.south) -- (response.north);

\draw[decorate, decoration={brace, amplitude=5pt, mirror}, gray!60] 
    ([xshift=-0.3cm]video.north west) -- ([xshift=-0.3cm]video.south west) 
    node[midway, left=6pt, timing] {parallel};

\end{tikzpicture}
\caption{ITAS spoke-and-wheel architecture. Three specialist agents execute in parallel; the synthesizer waits for all three before producing a response. End-to-end latency is dominated by $\max(L_v, L_g, L_c) + L_{\text{synth}}$.}
\label{fig:architecture}
\end{figure}
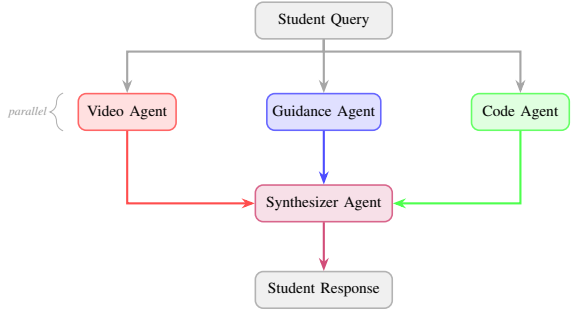

ITAS uses a spoke-and-wheel architecture (Fig.~\ref{fig:architecture}) built on Google's Agent Development Kit (ADK)~\cite{googleadk}. A student query is processed by four agents. Three \emph{specialist agents} execute in parallel:

\begin{itemize}[leftmargin=*]
    \item \textbf{Video Agent:} Receives the student query along with a transcript of the current lecture video segment. Connects the question to relevant lecture content and provides pedagogical context.
    \item \textbf{Code Agent:} Receives the student query along with the student's current code from the embedded IDE. Identifies bugs, suggests fixes, and explains relevant programming concepts.
    \item \textbf{Guidance Agent:} Receives the student query and the current lesson's learning objectives. Provides Socratic guidance without revealing solutions, aligned with the course pedagogy.
\end{itemize}

After all three parallel agents complete, a \textbf{Synthesizer Agent} receives their outputs and merges them into a single coherent response. All four agents use Gemini 2.5 Flash on Vertex AI with structured JSON output schemas to eliminate parsing overhead and with thinking disabled (\texttt{thinking\_budget=0}) to minimize latency.

\subsection{Latency Model}

We instrument the pipeline at five timing points:

\begin{itemize}[leftmargin=*]
    \item $T_0$: Request received at Cloud Run backend.
    \item $T_1$: Session state loaded and agent contexts constructed.
    \item $T_2$: Parallel agents dispatched.
    \item $T_3$: All parallel agents complete. $T_3 - T_2 = \max(L_v, L_c, L_g)$ where $L_v$, $L_c$, $L_g$ are the video, code, and guidance agent latencies.
    \item $T_4$: Synthesizer completes. $T_4 - T_3 = L_s$.
    \item $T_5$: Response returned to client.
\end{itemize}

End-to-end latency is $T_5 - T_0$. The critical observation is that the parallel phase $T_3 - T_2$ equals the \emph{maximum} of three independent LLM call latencies. The CDF $F(x)$ of a single agent transforms under the maximum operation to $F(x)^3$, shifting the distribution rightward. Concretely: taking the maximum of three parallel calls is like rolling three dice and keeping the highest; the expected value (4.96) exceeds the single-die mean (3.5), and the gap grows with variance. Throughput tiers that reduce per-agent variance therefore yield a \emph{disproportionate} benefit to multi-agent pipelines.

\subsection{Infrastructure}

ITAS runs on Google Cloud Run with auto-scaling. The Gemini 2.5 Flash API is accessed through Vertex AI in three throughput tiers:

\begin{itemize}[leftmargin=*]
    \item \textbf{Standard PayGo~\cite{standardpaygo}:} Pay-per-token with no reserved capacity. Requests join a shared inference pool alongside all other Vertex AI users. Pricing: \$0.30 per million input tokens, \$2.50 per million output tokens.
    \item \textbf{Priority PayGo~\cite{prioritypaygo}:} Pay-per-token at 1.8$\times$ standard rates, but requests are placed in a priority queue ahead of standard traffic. Uses Google's global endpoint. Pricing: \$0.54 per million input tokens, \$4.50 per million output tokens.
    \item \textbf{Provisioned Throughput~\cite{provisionedthroughput}:} Reserved capacity measured in Generative AI Scale Units (GSUs). Our benchmark allocation uses 7 GSUs, providing approximately 20,000 tokens per second of dedicated throughput. Pricing: \$2,700 per GSU per month at a 1-month commitment (\$18,900/month for 7 GSUs).
\end{itemize}

\section{Experiment Design}

\subsection{Live Deployment}

ITAS is deployed in a graduate STEM seminar at a research university. The benchmark corpus was extracted from a deployment that, at the time of corpus capture, covered the first 4 of 5 instructional modules; a fifth module was introduced later in the semester and is reported in the companion deployment paper~\cite{elhaimeur2026quantum}. This deployment runs on Standard PayGo Gemini 2.5 Flash in \texttt{us-east1} and provides the corpus of real student queries used in the controlled benchmark. Queries range from simple greetings to complex debugging questions involving domain-specific topics.

\subsection{Controlled Benchmark}

To isolate the effect of throughput tier and concurrency on latency, we extract approximately 100 real student interactions from the live seminar deployment, including chat messages and associated code context, and replay them under controlled conditions. Each of the three throughput tiers is tested at 11 concurrency levels (1, 5, 10, 15, 20, 25, 30, 35, 40, 45, 50 simultaneous users): Standard PayGo on \texttt{us-east1}, Priority PayGo on the \texttt{global} endpoint, and Provisioned Throughput (7 GSUs) on \texttt{us-central1}.

The benchmark uses a semaphore-based constant-concurrency model: exactly $N$ requests are in-flight at all times; as one completes, the next starts immediately. This models sustained classroom load more realistically than batch-wave approaches where all requests arrive simultaneously. Each request captures end-to-end latency, per-agent latency (video, code, guidance, synthesizer), parallel phase duration, per-agent input and output token counts, bottleneck agent identification, success/failure status, and traffic type confirmation (verifying which throughput tier actually served the request).

\subsection{Data Collection}

The benchmark produced over 3,000 total instrumented requests ($\sim$100 queries $\times$ 11 concurrency levels $\times$ 3 tiers). An \emph{error} is an HTTP 500 Internal Server Error from the Vertex AI inference endpoint, indicating the request was rejected or timed out. All three tiers maintained greater than 99\% success rates across all concurrency levels: Provisioned and Priority PayGo experienced zero errors out of 1,034 requests each; Standard PayGo experienced 2 errors (both at $c=20$) out of 1,034 requests.

\subsection{Limitations}
\label{sec:limitations}

The three throughput tiers use different Vertex AI endpoints: \texttt{global} for Priority PayGo, \texttt{us-central1} for Provisioned, and \texttt{us-east1} for Standard. Baseline latency differences at $c=1$ (2.8s vs.\ 3.7s vs.\ 4.1s) include a regional component. However, network round-trip differences are on the order of milliseconds, not seconds; we attribute the \emph{scaling behavior} across concurrency levels primarily to the throughput tier. All agents use thinking-disabled mode (\texttt{thinking\_budget=0}), which substantially reduces latency compared to default settings. All queries originate from a single graduate STEM seminar; query complexity may not represent all STEM courses, though the range includes simple greetings, conceptual questions, and complex debugging requests.

While our evaluation uses Google Vertex AI, the observed latency behaviors are not Vertex-specific artifacts. The distinction between shared inference pools, priority queues, and reserved capacity is common across cloud LLM providers; AWS Bedrock, Azure OpenAI Service, and Anthropic's API offerings expose structurally identical tier models. The parallel-phase maximum effect arises from queueing dynamics and inference variance that are properties of any shared or reserved compute pool, not of any single provider's implementation. We therefore expect the tier-selection findings to generalize to multi-agent pipelines deployed on other platforms, with the specific crossover concurrency levels varying according to each provider's pool characteristics and pricing.

\section{Performance Evaluation}

\subsection{End-to-End Latency}

\begin{figure}[t]
    \centering
    \includegraphics[width=\columnwidth]{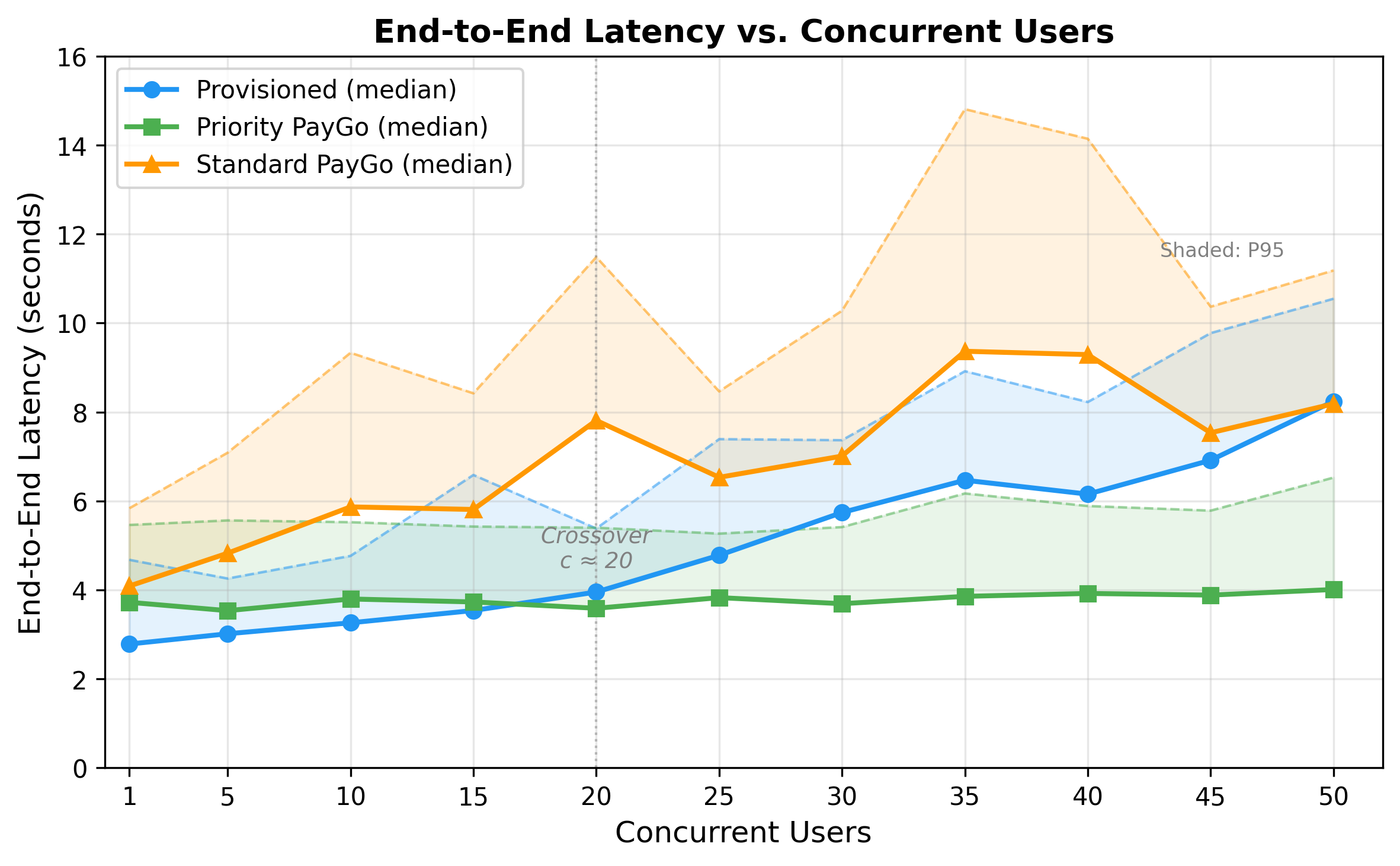}
    \caption{Median end-to-end latency and P95 band vs.\ concurrent users. Priority PayGo (green) maintains flat 3.5--4.0s median with a narrow P95 band throughout, consistent with low per-agent variance stabilizing the parallel-phase maximum. Provisioned (blue) starts lowest but its P95 diverges after $c=20$ as GSU saturation appears to introduce queuing variance. Standard (orange) shows the widest P95 band; its P95 reaches 14.1s at $c=40$.}
    \label{fig:latency_curves}
\end{figure}

Table~\ref{tab:latency} presents end-to-end latency across concurrency levels for all three tiers. Fig.~\ref{fig:latency_curves} visualizes the divergence.

\begin{table}[t]
\centering
\footnotesize
\caption{End-to-End Latency (seconds) by Throughput Tier}
\label{tab:latency}
\setlength{\tabcolsep}{4pt}
\begin{tabular}{@{}c
S S
S S S S
c@{}}
\toprule
& \multicolumn{2}{c}{\textbf{Reserved}} 
& \multicolumn{4}{c}{\textbf{On-Demand}} \\
\cmidrule(lr){2-3} \cmidrule(lr){4-7}
& \multicolumn{2}{c}{\textbf{Provisioned}} 
& \multicolumn{2}{c}{\textbf{Priority}} 
& \multicolumn{2}{c}{\textbf{Standard}} \\
\cmidrule(lr){2-3} \cmidrule(lr){4-5} \cmidrule(lr){6-7}
\textbf{$c$} & {Med} & {P95} & {Med} & {P95} & {Med} & {P95} & \textbf{Best} \\
\midrule
1  & 2.8 & 4.7 & 3.7 & 5.5 & 4.1 & 5.8 & Prov. \\
5  & 3.0 & 4.3 & 3.5 & 5.6 & 4.8 & 7.1 & Prov. \\
10 & 3.3 & 4.8 & 3.8 & 5.5 & 5.9 & 9.3 & Prov. \\
20 & 4.0 & 5.4 & 3.6 & 5.4 & 7.8 & 11.5 & Pri. \\
30 & 5.7 & 7.4 & 3.7 & 5.4 & 7.0 & 10.3 & Pri. \\
40 & 6.2 & 8.2 & 3.9 & 5.9 & 9.3 & 14.1 & Pri. \\
50 & 8.2 & 10.5 & 4.0 & 6.5 & 8.2 & 11.2 & Pri. \\
\bottomrule
\end{tabular}
\end{table}

The central finding is that Priority PayGo maintains remarkably flat latency: median stays within 3.5--4.0s from $c=1$ to $c=50$, a range spanning a 50$\times$ increase in load. Standard PayGo degrades from 4.1s to 9.3s, a 2.3$\times$ increase. Provisioned Throughput starts fastest at 2.8s but triples to 8.2s at $c=50$ as its 7-GSU allocation saturates.

The P95 bands reveal a deeper pattern. Priority PayGo's P95 band stays narrow throughout (P95 within 1.5--2.5s of median). This is consistent with a hypothesis that priority scheduling reduces tail latency by isolating priority traffic from shared-pool contention; we cannot rule out that regional or routing differences also contribute. Standard PayGo's P95 is wide and volatile, reaching 14.1s at $c=40$, more than double the median, consistent with unpredictable queuing in a shared inference pool under contention. Provisioned's P95 diverges from its median above $c=20$, suggesting the fixed GSU allocation begins queuing at that point. Whatever the mechanism, the variance behavior is the operative link to the parallel-phase maximum effect: a tier that stabilizes P95 stabilizes the maximum of three parallel draws from that distribution.

A crossover occurs at approximately $c=20$: below this point, Provisioned has the lowest latency, consistent with dedicated capacity and no queuing. Above $c=20$, Priority PayGo outperforms Provisioned. We attribute this to Priority PayGo drawing from a global priority pool with substantially more headroom than the fixed 7-GSU allocation, though we note this interpretation is consistent with, but not directly proven by, our measurements, as we did not instrument the provider-side queue. The crossover point is a property of our specific GSU allocation; provisioning more GSUs would shift it higher at proportionally higher cost.

\subsection{Pipeline Decomposition}

Table~\ref{tab:agents} decomposes latency across the four agent stages.

\begin{table}[t]
\centering
\footnotesize
\caption{Per-Agent Median Latency (s)}
\label{tab:agents}
\setlength{\tabcolsep}{4pt}
\begin{tabular}{@{}l c
S S S
S
S@{}}
\toprule
& & \multicolumn{3}{c}{\textbf{Parallel Agents}} & & \\
\cmidrule(lr){3-5}
\textbf{Tier} & \textbf{$c$} & {Video} & {Guid.} & {Code} & {Synth.} & {E2E} \\
\midrule
\multicolumn{7}{l}{\textit{Provisioned}} \\
\midrule
& 1  & 1.6 & 1.5 & 1.4 & 0.9 & 2.8 \\
& 10 & 2.0 & 1.8 & 1.6 & 1.1 & 3.3 \\
& 30 & 3.4 & 2.9 & 2.9 & 1.8 & 5.7 \\
& 50 & 4.7 & 4.4 & 4.4 & 2.4 & 8.2 \\
\midrule
\multicolumn{7}{l}{\textit{Priority PayGo}} \\
\midrule
& 1  & 2.1 & 1.8 & 1.8 & 1.3 & 3.7 \\
& 10 & 2.2 & 1.7 & 1.7 & 1.3 & 3.8 \\
& 30 & 2.2 & 1.9 & 1.8 & 1.3 & 3.7 \\
& 50 & 2.4 & 2.2 & 2.1 & 1.4 & 4.0 \\
\midrule
\multicolumn{7}{l}{\textit{Standard PayGo}} \\
\midrule
& 1  & 2.3 & 2.0 & 1.8 & 1.4 & 4.1 \\
& 10 & 3.0 & 2.9 & 2.9 & 2.1 & 5.9 \\
& 30 & 4.0 & 3.2 & 3.3 & 2.6 & 7.0 \\
& 50 & 4.4 & 3.9 & 3.7 & 3.1 & 8.2 \\
\bottomrule
\end{tabular}
\end{table}

The video agent is the bottleneck, the slowest of the three parallel agents, in 50--54\% of requests across all tiers, with guidance at 24--26\% and code at 21--24\%. This distribution is nearly identical across all three tiers and all concurrency levels, indicating the bottleneck is determined by input size, as the video agent receives the largest context (lecture transcript segments), and cannot be addressed by changing infrastructure.

The key contrast is in how per-agent latency scales. Priority PayGo's individual agents remain within 1.7--2.4s across the full concurrency range, consistent with the priority queue absorbing load without per-agent queuing pressure. Standard PayGo agents double (1.8--2.3s to 3.7--4.4s) and Provisioned agents triple (1.4--1.6s to 4.4--4.7s) as their respective pools saturate.

\subsection{The Parallel Max Effect}
\label{sec:parallelmax}

The parallel phase dominates end-to-end latency, accounting for 65--70\% of total time across all tiers. Because the system waits for the slowest of three agents, the parallel-phase duration exceeds the median of any individual agent. The magnitude of this inflation depends on the variance of per-agent latency.

At $c=10$, Standard PayGo individual agents have medians of 2.9--3.0s, but the parallel-phase median is 4.0s, a 37\% inflation. Priority PayGo individual medians of 1.7--2.2s yield a parallel-phase median of 2.4s, a 27\% inflation. The tighter the per-agent distribution, the smaller the penalty from taking the maximum. Latency optimization in multi-agent pipelines is therefore a problem of \emph{variance minimization}, not mean minimization, a distinction that does not arise in single-agent systems.

\subsection{Effective Throughput}

Table~\ref{tab:throughput} presents effective throughput, successful requests per minute, as a function of concurrency, along with \textbf{Conc/\textcent}: the number of concurrent users supported per cent of spend per minute. Priority PayGo scales approximately linearly to 748 requests per minute at $c=50$, while Provisioned plateaus near 390 and Standard near 370. At $c=50$, Priority PayGo delivers 2.0$\times$ the effective throughput of either alternative. Critically, the throughput plateau in Provisioned near $c=20$ directly corresponds to the latency crossover in Fig.~\ref{fig:latency_curves}: both are consistent with the same GSU saturation event, suggesting that reserved capacity and on-demand priority capacity diverge at the same load point from both a latency and throughput perspective.

\begin{figure}[t]
    \centering
    \includegraphics[width=\columnwidth]{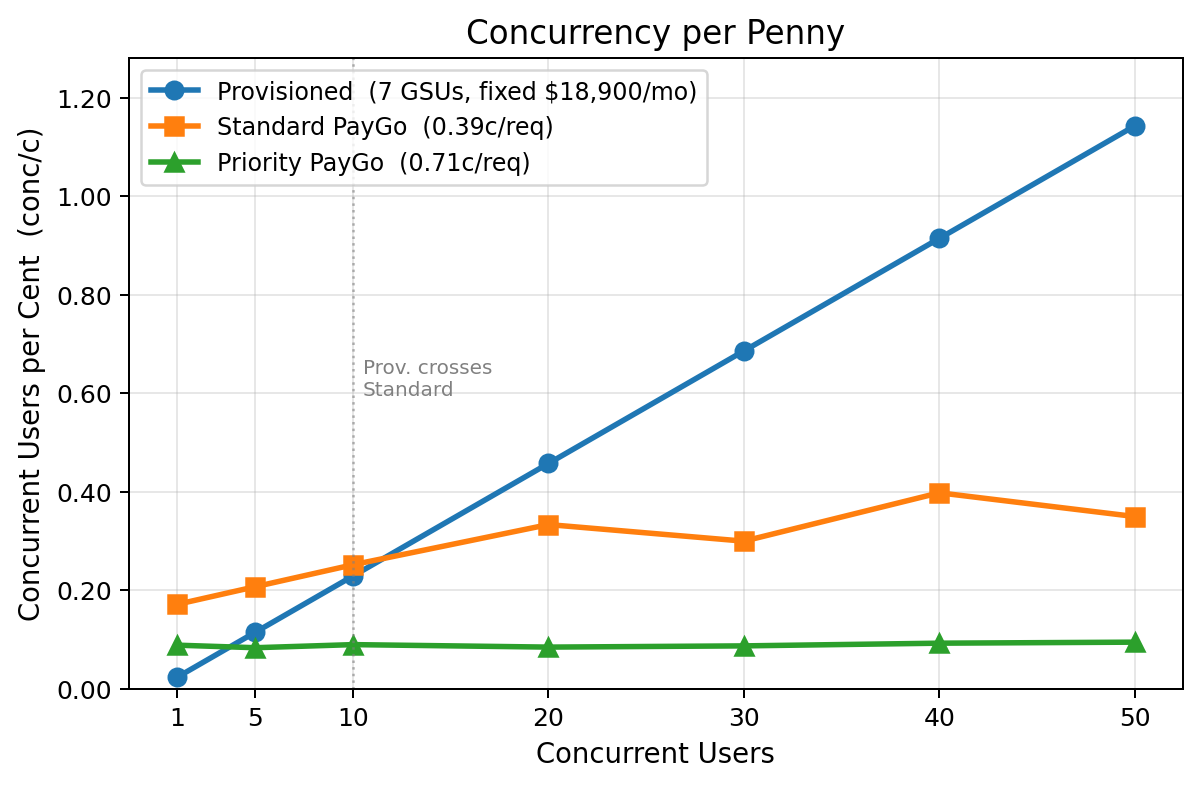}
    \caption{Concurrency per Penny (Conc/\textcent): concurrent users per cent of spend per minute. Provisioned (blue) starts low at $c=1$ but rises as the fixed cost amortizes across more users, crossing Standard near $c=10$. Standard PayGo (orange) leads among pay-per-token tiers. Priority PayGo (green) is flat near 0.09; its 1.8$\times$ premium is the cost of flat latency.}
    \label{fig:conc_per_penny}
\end{figure}

Conc/\textcent\ is computed as $c \div \text{cost\_per\_min}$ (in cents). For pay-per-token tiers, cost per minute scales with throughput: $\text{cost\_per\_min} = \text{rpm} \times C_{\text{tier}} \times 100$. For Provisioned, cost per minute is the fixed monthly rate amortized to 43.75\textcent/min regardless of utilization. Standard PayGo ($0.39$\textcent/req) delivers the highest Conc/\textcent\ among pay-per-token tiers, ranging from 0.17 to 0.40 conc/\textcent\ across concurrency levels. Priority PayGo ($0.71$\textcent/req) remains flat near 0.09 conc/\textcent; the 1.8$\times$ per-token premium directly reduces cost efficiency. Provisioned begins at only 0.02 conc/\textcent\ at $c=1$ (idle capacity) but rises to 1.14 at $c=50$ as the fixed cost is spread across more users. It crosses Standard near $c=10$ and continues to improve as utilization increases. Fig.~\ref{fig:conc_per_penny} visualizes these trajectories.

\begin{table}[t]
\centering
\footnotesize
\caption{Throughput and Cost Efficiency}
\label{tab:throughput}
\setlength{\tabcolsep}{4pt}
\begin{tabular}{@{}c
S S
S S S S@{}}
\toprule
& \multicolumn{2}{c}{\textbf{Reserved}} 
& \multicolumn{4}{c}{\textbf{On-Demand}} \\
\cmidrule(lr){2-3} \cmidrule(lr){4-7}
& \multicolumn{2}{c}{\textbf{Provisioned}} 
& \multicolumn{2}{c}{\textbf{Priority}} 
& \multicolumn{2}{c}{\textbf{Standard}} \\
\cmidrule(lr){2-3} \cmidrule(lr){4-5} \cmidrule(lr){6-7}
\textbf{$c$} & {Req/min} & {Conc/¢} & {Req/min} & {Conc/¢} & {Req/min} & {Conc/¢} \\
\midrule
1   & 22  & 0.02 & 16  & 0.09 & 15  & 0.17 \\
5   & 100 & 0.11 & 85  & 0.08 & 62  & 0.21 \\
10  & 184 & 0.23 & 158 & 0.09 & 102 & 0.25 \\
20  & 304 & 0.46 & 335 & 0.08 & 154 & 0.33 \\
30  & 314 & 0.69 & 488 & 0.09 & 257 & 0.30 \\
40  & 390 & 0.91 & 612 & 0.09 & 258 & 0.40 \\
50  & 364 & 1.14 & 748 & 0.09 & 367 & 0.35 \\
\bottomrule
\end{tabular}
\end{table}

\begin{figure}[t]
    \centering
    \includegraphics[width=\columnwidth]{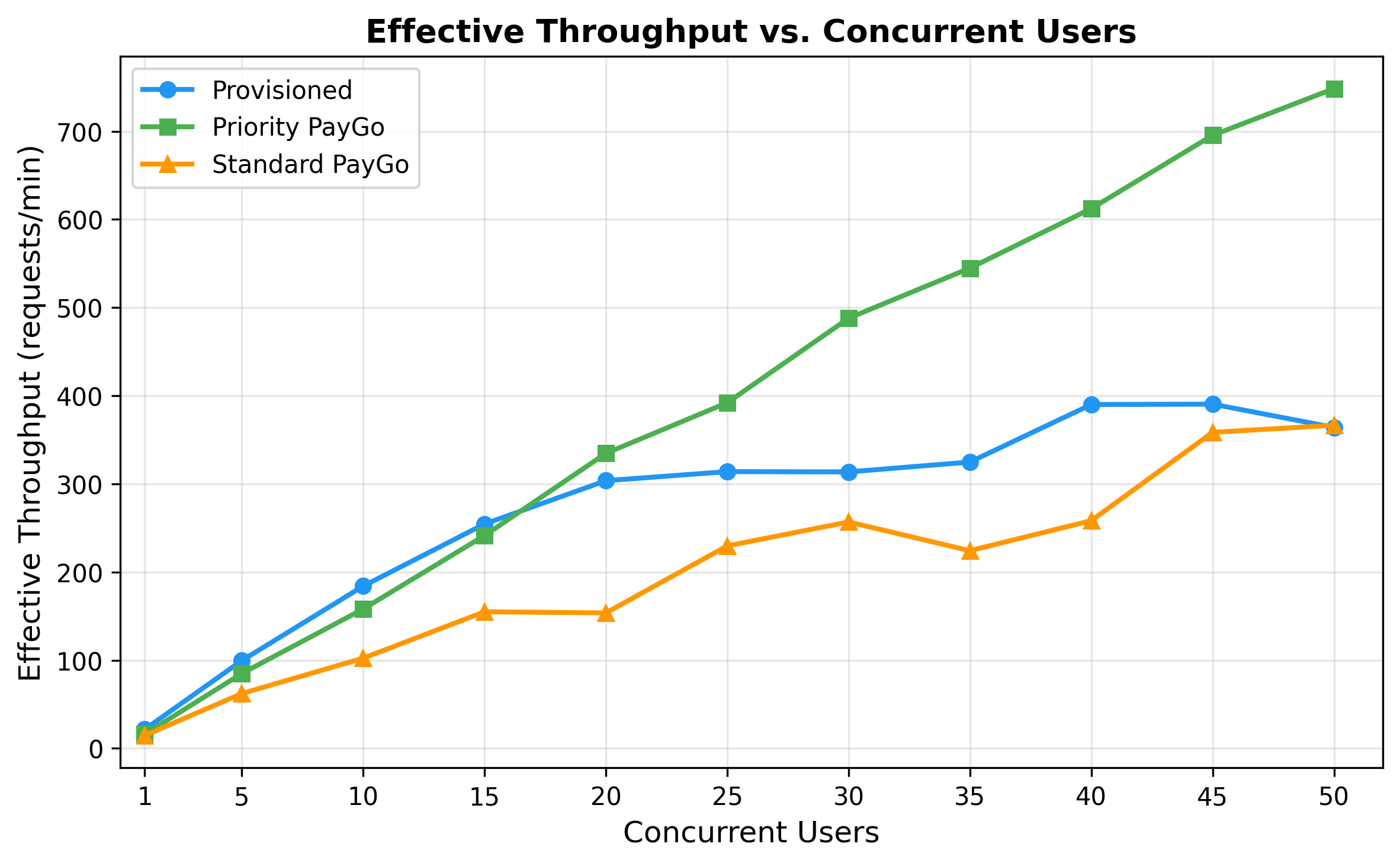}
    \caption{Effective throughput (req/min) vs.\ concurrent users. Priority PayGo (green) scales near-linearly to 748~req/min at $c=50$, consistent with no head-of-line blocking in the priority queue. Provisioned (blue) plateaus near 390~req/min at $c\approx20$, corresponding to the GSU saturation point. Standard (orange) shows irregular growth under shared pool contention.}
    \label{fig:throughput}
\end{figure}

\subsection{BiDi On-Demand: A Candidate Replacement for the Video Agent}
\label{sec:bidi}

The Video Agent is the bottleneck in 50--54\% of requests and the single agent whose removal would most reduce the parallel-phase maximum. BiDi provides a contrasting design point: rather than optimizing within the parallel fan-out architecture, it removes the parallel-phase maximum entirely for the video path by handling the response in a single streaming turn. The tradeoff is that BiDi operates as a standalone voice turn, losing the multi-agent synthesis that combines video context, code, and guidance in a single response.

\textbf{Model and tier constraints.} BiDi uses \texttt{gemini-live-2.5-flash-native-audio}, not the Gemini 2.5 Flash text model used by ITAS. The Live API offers only Standard (on-demand) and Provisioned tiers; Priority PayGo is unavailable. Our benchmark covers Standard BiDi only; a Provisioned configuration was not available at evaluation time.

\textbf{Pricing.} The Live API bills by modality: \$0.50/M input text tokens, \$3/M input audio, \$2/M output text, \$12/M output audio. Our benchmark used text queries, so cost estimates reflect text-input pricing and are a \emph{lower bound}; production audio input at \$3/M would raise cost roughly 6$\times$.

\textbf{Benchmark design.} Each session sends one query, records time to first audio byte (\texttt{first\_audio\_ms}) and total session duration (\texttt{total\_ms}), then interrupts immediately after the first response. No output audio was generated before interruption, so output audio cost is not captured.

Table~\ref{tab:bidi} reports median latency and estimated cost by concurrency level.

\begin{table}[t]
\centering
\footnotesize
\caption{BiDi On-Demand vs.\ ITAS Standard PayGo}
\label{tab:bidi}
\setlength{\tabcolsep}{3pt}
\begin{tabular}{@{}c
S[table-format=2.2]
S[table-format=2.2]
S[scientific-notation=true, table-format=1.1e-1]
S[table-format=2.1]@{}}
\toprule
\textbf{$c$} 
& \textbf{First Audio} 
& \textbf{BiDi E2E} 
& \textbf{Cost} 
& \textbf{ITAS E2E} \\
\midrule
1  & 0.79 & 2.75 & 3.4e-5 & 4.1 \\
5  & 0.83 & 2.71 & 3.5e-5 & 4.8 \\
10 & 0.96 & 2.90 & 3.8e-5 & 5.9 \\
20 & 1.48 & 3.88 & 5.1e-5 & 7.8 \\
30 & 2.23 & 5.79 & 7.0e-5 & 7.0 \\
40 & 2.58 & 5.94 & 7.9e-5 & 9.3 \\
50 & 3.32 & 7.56 & 1.0e-4 & 8.2 \\
\bottomrule
\end{tabular}
\end{table}

Two findings stand out. First, BiDi delivers \emph{sub-1s first-audio latency} for $c \leq 10$ (0.79s at $c=1$): the student hears the first word nearly instantly, compared to ITAS's 4.1s text delivery at the same concurrency. This is a qualitatively different user experience for voice-based tutoring. Second, BiDi input-text cost is roughly 60--115$\times$ lower per session than ITAS Standard PayGo, even before accounting for the multi-agent overhead in ITAS.

The limitation is scaling behavior. BiDi Standard (on-demand) uses the same shared inference pool as ITAS Standard PayGo, and its E2E latency degrades comparably: 2.75s at $c=1$ to 7.56s at $c=50$. First-audio latency exceeds 2s above $c=25$. Because Priority PayGo is unavailable for the Live API, there is no current Vertex AI mechanism to obtain Priority-tier latency stability for BiDi at classroom scale.

A Provisioned BiDi configuration, which we did not benchmark, is the natural next step: it would provide dedicated capacity for the Live API, potentially maintaining sub-1s first-audio latency at higher concurrency. This remains an open question for future work. From a system design perspective, replacing the Video Agent with a well-provisioned BiDi configuration would eliminate the most frequent bottleneck from the parallel phase entirely, potentially reducing the parallel-phase maximum from $\sim$2.1s (Priority PayGo Video Agent at $c=1$) to BiDi's first-audio latency of $\sim$0.8s, a 60\% reduction in the dominant latency term and a direct application of the variance minimization principle identified in \S\ref{sec:parallelmax}.

\section{Cost Analysis}

\subsection{Per-Request Cost}

Token usage is stable across concurrency levels (concurrency affects queuing, not content). Across all 3,100 successful requests, mean token usage is 6,671 input tokens and 767 output tokens per request, totaling 7,438 tokens across all four agents.

Using Vertex AI pricing for Gemini 2.5 Flash (non-thinking mode):

\bigskip
\textbf{Standard PayGo} (\$0.30/M input, \$2.50/M output):
\begin{equation}
    C_{\text{std}} = \frac{6{,}671 \times 0.30 + 767 \times 2.50}{10^6} \approx \$0.0039
\end{equation}

\bigskip

\textbf{Priority PayGo} (\$0.54/M input, \$4.50/M output):
\begin{equation}
    C_{\text{pri}} = \frac{6{,}671 \times 0.54 + 767 \times 4.50}{10^6} \approx \$0.0071
\end{equation}
\medskip

Priority PayGo costs 1.8$\times$ Standard per request, the price of flat latency at any concurrency level.

\subsection{Provisioned Throughput Pricing}

Provisioned capacity is purchased at a fixed rate per GSU independent of actual usage. Rates range from \$2,700/GSU/month (1-month commitment) to \$2,000/GSU/month (12-month commitment). Our benchmark uses 7 GSUs at the 1-month rate: \$18,900/month providing approximately 20,000 tokens per second of dedicated throughput.

\subsection{Per-Student Cost at Scale}

We stress-test the cost model at 10,000 questions per student per semester, meaning 100 questions per day, every day, for a full semester. This is a deliberate upper bound: it abstracts away real-world usage patterns such as uneven question rates, idle periods between lectures, and query volume variation across students. No realistic student usage pattern would approach this volume; actual per-student costs will be substantially lower. To ground the estimate, consider a more plausible scenario: 15 questions per day on class days (roughly 45 days of instruction per semester), totaling 675 questions. At that rate, Priority PayGo costs $675 \times \$0.0071 \approx \$4.79$ per student, less than a single textbook chapter. Standard PayGo costs $\approx \$2.63$. The 10,000-question ceiling in Table~\ref{tab:cost_scale} should therefore be read as a worst-case bound, not a predicted cost.

\begin{table}[t]
\centering
\footnotesize
\caption{Per-Student Cost per Semester (10,000 queries)}
\label{tab:cost_scale}
\setlength{\tabcolsep}{5pt}
\begin{tabular}{@{}l S S S S@{}}
\toprule
\textbf{Scale} & {Standard} & {Priority} & {Provisioned} & {Prov./Pri.} \\
\midrule
40 students   & 39  & 71  & 1890 & 27 \\
400 students  & 39  & 71  & 945  & 13 \\
4K students   & 39  & 71  & 473  & 7  \\
16K students  & 39  & 71  & 236  & 3  \\
\midrule
\textit{STEM textbook} & \multicolumn{4}{c}{\textit{$\sim$150 per student}} \\
\bottomrule
\end{tabular}
\end{table}

Standard PayGo at \$39.26 per student and Priority PayGo at \$70.67 per student are both well below the cost of a typical STEM textbook (\$100--200), even under this extreme usage assumption. These costs are flat regardless of the number of students because pay-per-token pricing scales linearly with usage. Unlike traditional ITS infrastructure costs, which scale with peak capacity regardless of actual demand, pay-per-token LLM systems incur cost only when students are actively querying, a fundamental shift in the cost model that makes per-student cost a stable, predictable quantity.

Provisioned Throughput has a fundamentally different cost structure governed by \emph{utilization}, not scale. The estimates in Table~\ref{tab:cost_scale} assume continuous 24/7 provisioning, the worst case, in which the reserved capacity sits idle outside of active use. Under that assumption, the per-student cost remains above \$225 at any enrollment. However, this is not an inherent property of Provisioned Throughput; it is a consequence of underutilization.

An institution that can \emph{predict or anticipate usage}, whether from historical query logs, class enrollment data, or known course schedules, can provision capacity only when demand is expected, releasing it otherwise. The goal is to pack the reserved window as close to 100\% utilization as possible: if GSUs are active only when students are actively querying, the effective cost per student drops proportionally. At full utilization, Fig.~\ref{fig:conc_per_penny} already shows Provisioned as the most cost-efficient tier. The critical variable is therefore not the size of the institution but the institution's ability to forecast and concentrate its traffic. Deployments that can achieve high utilization, through scheduling, usage-based provisioning windows, or traffic shaping, may find Provisioned Throughput the most economical choice. Those that cannot predict demand or require always-on availability will pay for idle capacity, making pay-per-token tiers the safer default.

\begin{figure}[t]
    \centering
    \includegraphics[width=\columnwidth]{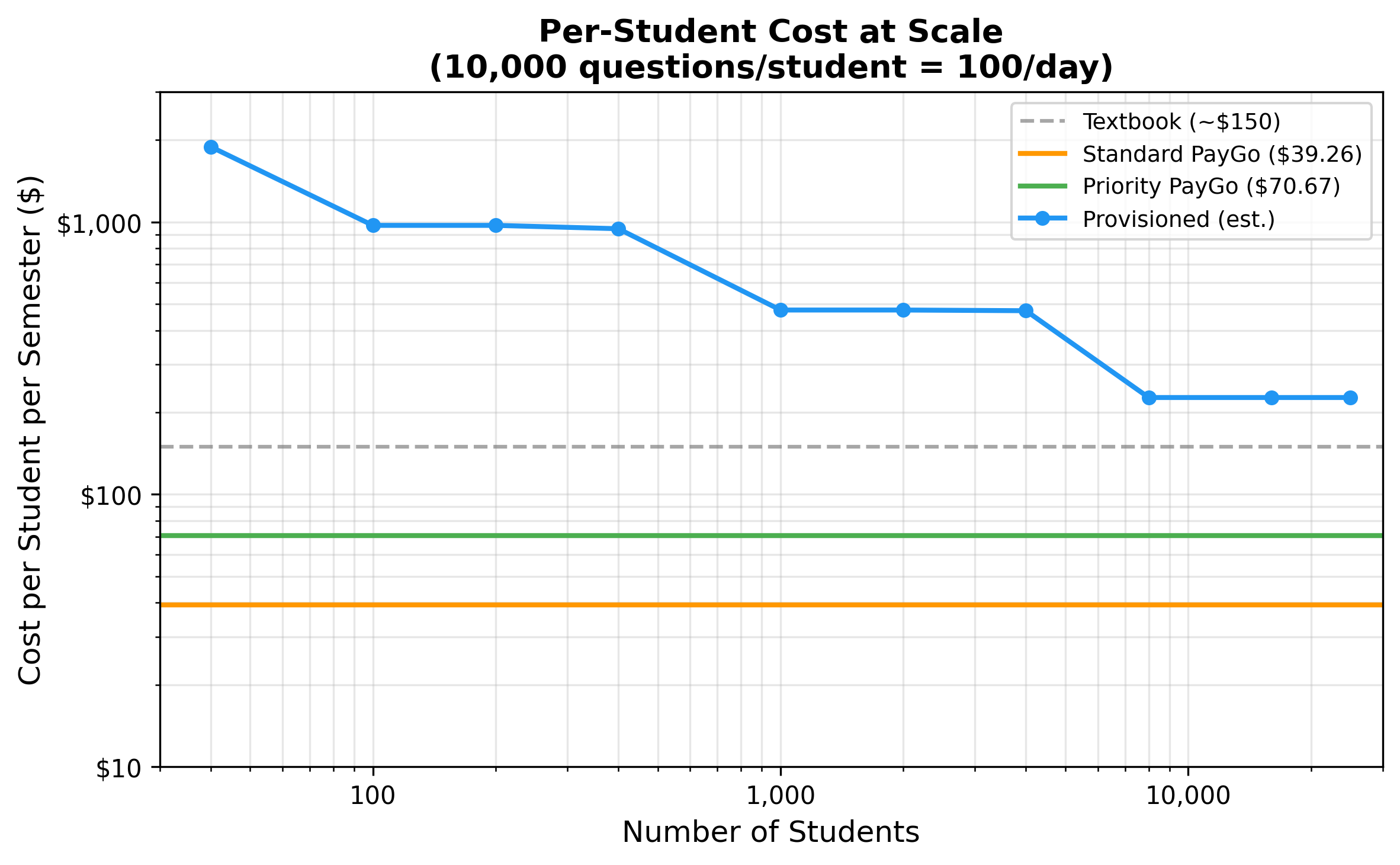}
    \caption{Per-student cost at scale (10,000 questions/semester, worst-case ceiling, 24/7 provisioning assumed). Standard and Priority PayGo are flat at \$39 and \$71. Provisioned (dashed) is an upper bound; institutions with predictable traffic will see substantially lower costs.}
    \label{fig:cost}
\end{figure}

Fig.~\ref{fig:cost} visualizes the continuous-provisioning upper bound across all tiers.

\section{Scalability Model}

Based on our measurements, we identify four scaling regimes (Table~\ref{tab:scaling}).

\begin{table}[t]
\centering
\footnotesize
\caption{Scaling Regimes and Recommendations}
\label{tab:scaling}
\setlength{\tabcolsep}{4pt}
\begin{tabular}{@{}l c c c c c@{}}
\toprule
\textbf{Scale} & \textbf{$n$} & \textbf{Peak $c$} & \textbf{Tier} & \textbf{Latency} & \textbf{Cost} \\
\midrule
Seminar      & 5--10   & $\leq$10   & Std. & 5.9s & \$39 \\
Classroom    & $\sim$40     & $\sim$20   & Pri. & 3.6s & \$71 \\
Department   & $\sim$400    & $\sim$100  & Pri. & $<$4s & \$71 \\
University   & 4K--25K & 500+        & Pri. & $<$4s & \$71 \\
\bottomrule
\end{tabular}
\end{table}

\textbf{Seminar (5--10 students, $c \leq 10$).} Standard PayGo is adequate and cost-optimal. Our live seminar deployment operates in this regime with median latency under 6s.

\textbf{Classroom (40 students, peak $c \approx 20$).} Priority PayGo is recommended. At $c=20$, Priority maintains 3.6s median while Standard degrades to 7.8s. The additional cost (\$71 vs.\ \$39 per student at extreme usage) is modest relative to the 2.2$\times$ latency improvement.

\textbf{Department (400 students, peak $c \approx 100$).} Priority PayGo requires no architectural changes; it scales transparently because Google manages the capacity. The only infrastructure concern is horizontal scaling of the Cloud Run application layer and session state management.

\textbf{University (4,000--25,000 students, peak $c \approx 500$--1,000+).} Priority PayGo continues to scale because it draws from Google's global priority pool. The primary architectural requirements are multi-region Cloud Run deployment for geographic distribution and distributed session state. For institutions with globally distributed students (e.g., distance-learning programs), geographic distribution actually \emph{reduces} peak concurrency by spreading usage across time zones.

\begin{quote}
\textbf{Remark:} Priority PayGo requires no capacity planning and scales elastically with demand at a fixed per-token rate, making it the recommended default for any deployment where concurrent usage exceeds 10 students and traffic is difficult to predict. Provisioned Throughput becomes competitive, and can be the most cost-efficient option, when an institution can accurately forecast demand and schedule reservations to maintain high utilization. The choice between them is ultimately a question of traffic predictability, not scale.
\end{quote}

Provisioned Throughput is a viable alternative for institutions that can anticipate their traffic. At high utilization it is the most cost-efficient tier and provides the lowest latency at low concurrency. The open question is operational: whether a given institution has the usage data and scheduling discipline to keep reserved capacity well-utilized.

\section{Discussion}

\textbf{Priority PayGo as the dominant tradeoff point.} Across all evaluated concurrency regimes, Priority PayGo is the best-performing throughput tier for multi-agent tutoring deployment. Its flat latency profile (3.5--4.0s regardless of concurrency), zero failure rate, absence of capacity planning, and per-student cost below a textbook make it the strongest choice for classroom-through-university scale among the tiers we measured. The 1.8$\times$ per-token premium over Standard PayGo is the cost of consistent sub-4-second response times, a worthwhile trade for interactive educational applications where response latency directly affects student engagement.

\textbf{The multi-agent tradeoff.} One could eliminate the parallel-phase overhead by collapsing to a single agent with a combined prompt. Our prior work~\cite{elhaimeur2025toward} shows that agent specialization improves response quality and pedagogical alignment. The latency cost of the multi-agent design is quantifiable from our data: on Priority PayGo at $c=1$, a single specialist agent completes in approximately 2.1s; the parallel-phase maximum adds roughly 0.6s over that, and the synthesizer adds a further 1.3s, for a total overhead of~$\sim$1.9s relative to a hypothetical single-agent call. Priority PayGo makes this tradeoff favorable: the overhead is stable and small in absolute terms, and the specialization benefit, demonstrated in prior work, justifies it. A single collapsed agent would save the overhead but lose the ability to simultaneously contextualize video content, inspect live code, and apply Socratic guidance in one response.

\textbf{Provisioned throughput: cost-effective only when utilization is predictable.} The cost model in Table~\ref{tab:cost_scale} assumes continuous 24/7 provisioning, a worst-case utilization assumption that makes Provisioned appear expensive at every scale. This framing is misleading for institutions that can predict their usage. University courses run on fixed schedules: lectures meet at known times, lab sessions have defined durations, and exam periods are foreseeable. An institution that provisions GSUs only during active class windows, say, 90 hours per month out of 720, pays roughly 12.5\% of the always-on cost, reducing the effective per-student expense proportionally. Fig.~\ref{fig:conc_per_penny} already captures this dynamic: as utilization rises toward 100\%, Provisioned's Conc/\textcent\ surpasses both pay-per-token tiers by a wide margin, reaching 1.14 conc/\textcent\ at $c=50$ against Standard's 0.35 and Priority's 0.09. The critical variable is therefore not scale, but \emph{how accurately an institution can forecast demand and pack the reserved capacity}. For deployments with unpredictable or bursty usage, pay-per-token tiers remain the safer default. For deployments with known, recurring traffic windows, Provisioned Throughput can be the most cost-efficient option of the three, while also delivering the lowest latency at low concurrency (2.8s vs.\ 3.7s at $c=1$) and providing independence from Google's shared inference pool.

\textbf{Generalizability beyond Vertex AI.} While this evaluation targets Google Vertex AI, the behaviors observed are not provider-specific. The fundamental distinction between shared inference pools, priority queues, and reserved capacity exists across AWS Bedrock, Azure OpenAI Service, and other cloud LLM providers under different product names. The parallel-phase maximum effect is a property of any architecture that fans out concurrent API calls to a shared or tiered inference backend; the specific crossover concurrency and cost thresholds will differ by provider, model, and pricing structure, but the qualitative pattern should hold broadly: reserved capacity wins at low concurrency, elastic priority capacity wins at high concurrency, and both beat shared-pool at scale.

\textbf{Quality of service.} This paper focuses on latency and cost rather than pedagogical quality; a rigorous evaluation of response quality is beyond its scope. We note informally that across the throughput tiers evaluated, responses sampled during the live deployment showed no perceptible degradation in quality as the system was moved between tiers, as answer accuracy, code suggestions, and explanatory depth appeared consistent regardless of whether requests were served by Standard, Priority, or Provisioned capacity. This is expected: throughput tiers affect \emph{when} a request is served, not \emph{how} the model processes it; the underlying model weights and inference are identical across tiers. We report this observation without statistical rigor and include it only to note that the latency-cost tradeoffs studied here do not appear to introduce a quality penalty. Formal quality evaluation comparing ITAS responses to expert instruction remains future work.

\section{Future Work}

Several optimizations could further reduce ITAS latency from the current 3.5s Priority PayGo baseline. \textbf{Context caching}, supported by Vertex AI for Gemini models, would cache the static lesson instructions per agent rather than reprocessing them per request, potentially reducing per-agent latency by 40--60\%. \textbf{Adaptive agent selection} would route simple queries to only the relevant agent(s), reducing the parallel phase to a single-agent call for straightforward questions. \textbf{Synthesizer elimination}, replacing the LLM-based merge with structured JSON composition, would remove approximately 1.3s of sequential overhead.

A \textbf{hybrid Priority PayGo and Provisioned} configuration, using provisioned capacity as a baseline with automatic spillover to Priority PayGo during peak demand, is natively supported by Vertex AI and could combine the low-concurrency latency advantage of provisioned throughput with Priority PayGo's elastic scaling.

Beyond optimization, \textbf{formal quality evaluation} comparing ITAS responses to expert teaching assistant responses would establish whether the system's pedagogical quality justifies deployment as a supplement to, or partial replacement for, human tutoring. A \textbf{formal controlled study} running both throughput tiers on the same class with random assignment would strengthen the causal argument that the benchmark results support. The bi-directional lecture format, where AI delivers lecture content with live student interaction, introduces streaming audio latency as an additional dimension.

\section{Conclusion}

Multi-agent tutoring architectures have the potential to deliver pedagogical benefits through specialization, but each student interaction generates multiple concurrent API calls whose latencies compound through the parallel-phase maximum. We have shown that Google Vertex AI's Priority PayGo tier uniquely mitigates this compound latency, maintaining flat 3.5--4.0s median response times from 1 to 50 concurrent users while Standard PayGo degrades to 9.3s and Provisioned Throughput, despite starting at the lowest latency of 2.8s, saturates its reserved capacity and crosses above Priority at $c \approx 20$.

The economic picture is straightforward. Even at an extreme usage assumption of 100 questions per day per student, Priority PayGo costs \$70.67 per student per semester, roughly half a STEM textbook. Standard PayGo costs \$39.26, roughly a quarter. Provisioned Throughput's cost under continuous 24/7 provisioning remains above \$225 per student at any scale, but this is a worst-case bound: institutions that can predict and concentrate their traffic can provision efficiently and achieve the highest cost efficiency of any tier.

The remaining open question at seminar and classroom scale is not cost or latency, as both are tractable with the tiers and configurations evaluated here. At larger scales, open engineering questions remain: capacity planning for Provisioned deployments with variable demand, multi-region session state, and the latency implications of context caching on the parallel max effect. The deeper open question across all scales is pedagogical quality: whether multi-agent AI tutoring can match the nuanced, adaptive guidance of expert human instruction. Our data suggests that the infrastructure investment is tractable at scale; the return depends on the quality of what the system teaches.

\section*{Acknowledgments}
This research was sponsored in part by the Richard T. Cheng Endowment and supported by Monarch Sphere~\cite{monarchsphere}. Cloud infrastructure was provided through Google Cloud Platform research credits and John D. Pratt, Seth J. Hohensee, and Alex L. Tucker of the ITS Group at Old Dominion University. The QIS curriculum is based on John Watrous's IBM Quantum lecture series. ITAS is developed at the Center for Real-Time Computing (CRTC), Old Dominion University. Gemini was used to improve readability across the article; the authors take full responsibility for all content.

\bibliographystyle{IEEEtran}
\bibliography{ref}

\end{document}